# Linear Structures in the Core of the Coma Cluster of Galaxies

J. S. Sanders[1,2*], A. C. Fabian[2], E. Churazov[3,4], A. A. Schekochihin[5], A. Simionescu[6,7,8], S. A. Walker[2], N. Werner[6,7]

[1]Max-Planck-Institut für extraterrestrische Physik, Giessenbachstrasse 1, 85748 Garching, Germany
[2]Institute of Astronomy, University of Cambridge, Madingley Road, Cambridge, CB3 0HA, UK
[3]Max-Planck-Institut für Astrophysik, Karl-Schwarzschild-Strasse 1, 85748 Garching, Germany
[4]Space Research Institute (IKI), Profsoyuznaya 84/32, Moscow 117997, Russia
[5]Rudolf Peierls Centre for Theoretical Physics, University of Oxford, 1 Keble Road, Oxford, OX1 3NP, UK
[6]KIPAC, Stanford University, 452 Lomita Mall, Stanford, CA 94305, USA
[7]Department of Physics, Stanford University, 382 Via Pueblo Mall, Stanford, CA 94305-4060, USA
[8]Institute of Space and Astronautical Science (ISAS), JAXA, 3-1-1 Yoshinodai, Chuo-ku, Sagamihara, Kanagawa 252-5210, Japan
*Correspondence to:  jsanders@mpe.mpg.de

**The hot X-ray emitting plasma in galaxy clusters is predicted to have turbulent motions which can contribute around ten percent of the cluster's central energy density. We report deep Chandra X-ray Observatory observations of the Coma cluster core, showing the presence of quasi-linear high-density arms spanning 150 kpc, consisting of low-entropy material likely stripped from merging subclusters. Two appear to be connected with a subgroup of galaxies at 650 kpc radius that is merging into the cluster, implying coherence over several hundred Myr. Such long lifetime implies that strong isotropic turbulence and conduction are suppressed in the core, despite the unrelaxed state of the cluster. Magnetic fields are presumably responsible. The structures seen in Coma present insight into the past Gyr of subcluster merger activity.**

Galaxy clusters are the largest gravitationally bound structures and are dominated by dark matter. They are the latest structures to form in the cosmological hierarchical structure formation scenario. They grow through mergers and by the accretion of matter. Most of their baryonic matter consists of a hot plasma (the intracluster medium, or ICM), heated during cluster formation to temperatures of several $10^7$ K, and visible by its X-ray emission. Cluster major mergers, the mergers between clusters of similar mass, are the most energetic events in the local universe, injecting turbulence and motions contributing to the total ICM energy density by around 10% of the thermal value (*1, 2*), increasing with radius.

The Coma cluster of galaxies is one of the best studied nearby rich clusters. However, it is not dynamically relaxed. X-ray observations have shown the disturbed nature of its ICM (*3, 4*). In particular, a group of galaxies associated with NGC 4839 to the south-west is merging with the main cluster. Unusually, the cluster has two central giant elliptical galaxies, NGC 4874 and NGC 4889, with a 700 km s$^{-1}$ line-of-sight velocity difference (*5*). The distribution of galaxy velocities within the cluster implies several different subgroups.

We examined the core of the Coma cluster (Fig. 1) with a 546-ks-deep set of Chandra observations (*6*). At the ~9 keV ($10^8$ K) temperature of the ICM in Coma, X-ray surface brightness variations are mainly due to density fluctuations.

The most striking features are a set of high surface brightness 'arms' (labeled A1 to A4). A1 is enhanced in surface brightness with respect to the larger-scale emission by up to around 10%, although this can increase to 15 or 20% using different models of the underlying surface brightness. Unless we are viewing these features from a special direction, the most likely morphology of the arms is that they are roughly cylindrical and are as deep as they are wide on the sky. We estimate that they are enhanced in density by 35–40%, assuming that they lie in the plane of the sky at the mid-point of the cluster. The fractional density enhancement relative to their surroundings increases if the arms are significantly in front or behind the cluster. In addition, if they are inclined along the line of sight, their surface brightness would be boosted by $1/\sin \theta$ where $\theta$ is the angle between the arm and the line of sight. If this is the case, the density in the arms would be smaller than our estimates. Small values of $\theta$ will boost their chances of being detected but their intrinsic lengths would be much larger than observed on the sky.

Around each of NGC 4889 and NGC 4874, are bright dense 'halos' of gas, labeled in the top panel H1 and H2, respectively. They are larger than the compact galactic mini-coronae (*7*), which we do not examine here. A more detailed view of the region between these galaxies is shown in Fig. 2. There is also a region of enhancement, labeled H3, around 100 kpc to the north, east and south of NGC 4874. An edge to this region between NGC 4889 and NGC 4874 is apparent (labeled E). This may be a shock (the projected pressure jumps by 10%), but the data are insufficient to verify the expected temperature change.

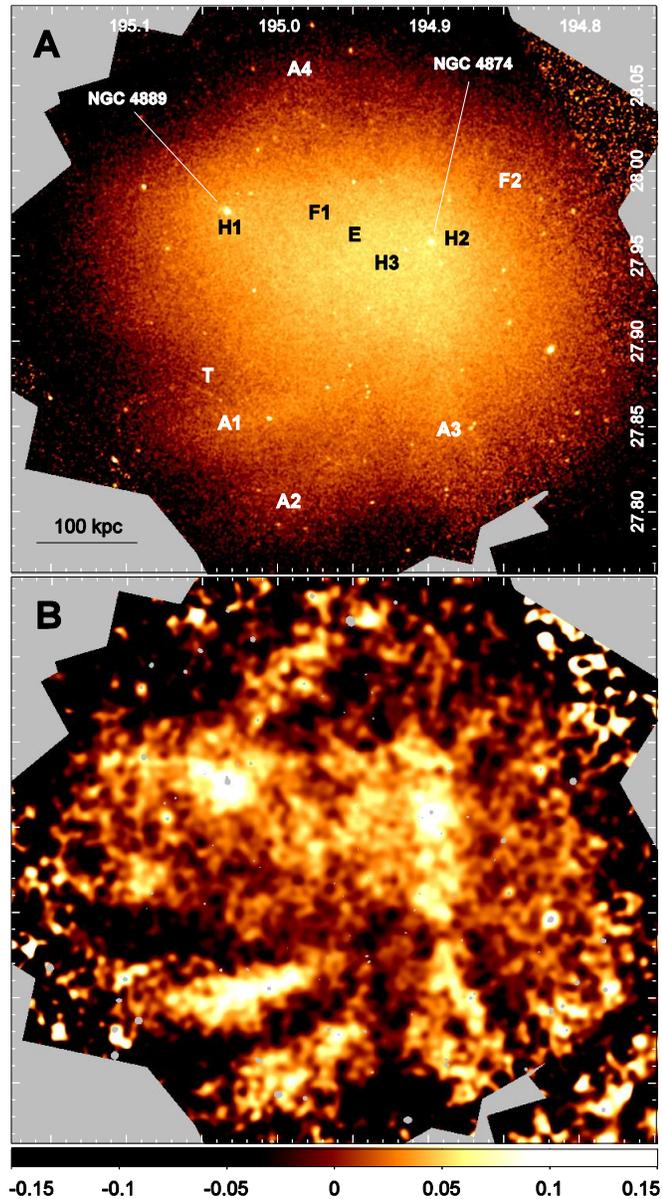

**Fig. 1**. Chandra X-ray image of Coma. (**A**) X-ray surface brightness with labeled features. Coordinates are J2000 decimal degrees. (**B**) Unsharp-masked image of the same region, enhancing features below ~30 kpc. The numerical values show the fractional surface brightness enhancement or suppression in smaller-scale features relative to larger-scale emission. Grey areas lie outside the dataset or are excluded point sources.

Between this edge and NGC 4889 are 50-kpc-long filamentary features running in the north-east to south-west direction (marked as F1) and also radiating away from NGC 4874 to the north-west and west (labeled F2). These filamentary features are enhanced by approximately 5% in surface brightness and are not associated with detector features (Fig. S1). The minimum length scale observed in the X-ray image is around the mean free path for electrons here (~7 kpc). These linear structures are likely to require magnetic fields in order to be stable.

Behind the galaxy GMP 2910 there is a remarkably straight 48 kpc long soft X-ray tail (labeled T in Fig. 1A). The spectrum of this material is consistent with thermal material at ~1 keV temperature (Fig. S2). This tail of presumably stripped gas was also seen by its Hα emission (8). A similar tail of stripped gas with X-ray emission was also seen in A3627 (9).

There is a gradient in emission-weighted projected temperature from the hot north-west side of the core to the cool south-east, previously seen using ASCA (10) and XMM (11) (Fig. 3). The arms are cooler than their surroundings and the gaps between the arms. There is also a cooler region north and south of NGC 4874, along the direction of arm A3. Accounting for projection effects, the pressure enhancement of A1 and A3 relative to their surroundings is 18±13 and 24±11%, statistically, respectively. However, if the arms are not in the plane of the sky or if our geometrical assumptions are incorrect, the above pressure values will be reduced. Therefore the arms are consistent with being in pressure equilibrium with their surroundings.

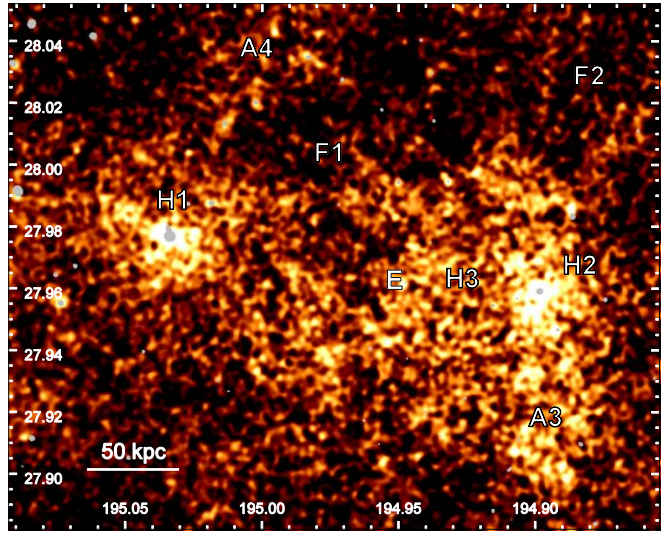

**Fig. 2.** Detailed view of the cluster center. A more detailed unsharp-masked image of the region around the two central galaxies.

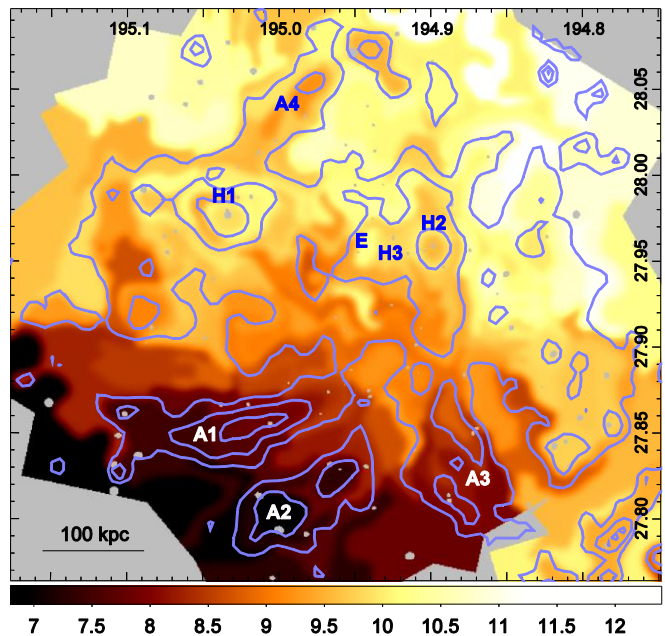

**Fig. 3.** Temperature map. Emission-weighted ICM temperature (keV), calculated by spectral fitting. The contours show lines of constant surface brightness enhancement in an unsharp-masked map with levels of 0, 3.5 and 7%.

Arm A1 has a 'specific entropy' 30% lower than its surroundings, assuming a cylindrical morphology in the plane of the sky. In a relaxed galaxy cluster the lowest entropy material lies at the bottom of the potential well. In Coma, the lowest entropy X-ray emitting material appears to lie in the arms, offset from the core itself. We estimate that arm A1 contains around 4×10¹⁰ Solar masses of material.

To examine the connection of the observed structures to those on larger scales, we examined XMM-Newton observations of the fractional deviation in the surface brightness from the average at each radius (Fig. 4). Arms A1 and A2 are immediately apparent, and A3 and A4 are also visible. The strongest feature (labeled L1) is a bright arm of emission extending from the center towards NGC 4911 and then towards NGC 4921, seen previously using ROSAT (*12*) and XMM (*13*). NGC 4911 is likely part of a distinct high velocity group merging with the main cluster (*5*). L1 may turn back towards the centre of the cluster as L2, towards the central galaxy NGC 4889.

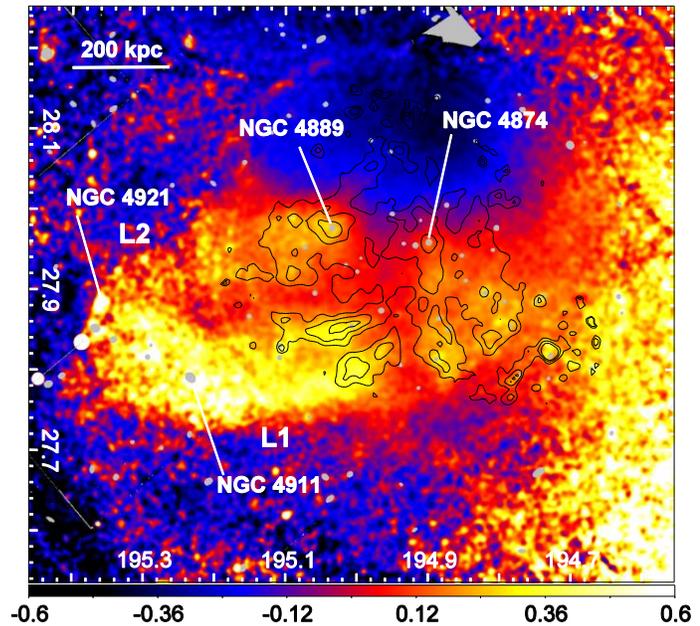

**Fig. 4.** Larger scale X-ray distribution. XMM EPIC-MOS image of the fractional difference in surface brightness from the average at each radius. The two halves of the large scale structure, L1 and L2, are labeled. The contours are as in Fig. 3.

A2 appears to be the extension of L1 into the core. A1 may also be part of this structure, if L1 bifurcates, or may be a separate feature. The whole L1 structure spans from 200 kpc radius in the centre of the cluster out to radii of at least 650 kpc, or further if the feature is not in the plane of the sky. The central galaxy NGC 4889 itself may also be part of this large-scale structure.

The arms and other features in the core of Coma are likely to be the result of the gas stripped from subclusters merging with the cluster. The stripped gas retains memory of the temperature and entropy of the subcluster. The lower entropy of the arms relative to their surroundings supports the hypothesis that they are stripped material. The tail of stripped material behind GMP 2910 (labeled T in Fig. 1A) demonstrates that stripping is occurring in Coma. Stripping is clearly seen in the Virgo Cluster, where a subcluster of galaxies associated with M86 is merging with the main cluster (*14*). It has been highlighted (*15*) that low-entropy gas associated with the merging group can give rise to density variations. However, there appear to be several arms in Coma not associated with the NGC 4911 group.

If arms A1, A2 and L1 are stripped material from the NGC 4911 subgroup merger, then such material can remain coherent over several hundred kpc. The arms are likely to be around 300 Myr old (taking the projected radius of NGC 4911 and dividing by the sound speed). The linear morphology of the arms is suggestive of laminar flow, indicating a lack of strong turbulence and a relatively high effective viscosity. Measurements of surface brightness fluctuations also indicate few strong density variations (*15*). The arms, however, are massive and require timescales of ~70 Myr to be accelerated by ram pressure to the sound speed and so do not instantaneously trace the ICM motion. Isotropic conduction with the surroundings must be suppressed, likely by magnetic fields. The timescale for conduction at the Spitzer level to equilibrate the temperature is around 8 Myr, reducing if the length of the conduction zone is shorter than the arm width. This is much shorter than their 300 Myr age, assuming that they are due to stripping from the NGC 4911 subgroup. Tangential motion of structures through a cluster can induce magnetic draping (*16*), where field lines are stretched, suppressing transport processes with the surroundings.

The NGC 4911 subgroup could have passed through the cluster core from the west to the east, leaving stripped material A1 and A2 in its wake. L1 is also cooler than its surroundings (*11, 17*), implying that it has a different origin from the bulk cluster gas. Arm A3 points roughly in the direction of the NGC 4839 group, so could be related to this merger. This supports the argument that the group has already passed through the cluster center (*18*), because the structure could not precede the galaxies in the group. However, it should be noted that A3 appears to end close to the edge of the Chandra field (Fig. 4) and does not continue any further towards NGC 4839. There is also a chain of cluster member galaxies along the direction of A3 (*5*) and a velocity substructure is detected nearby and also close to the inner part of A2 (*19*). The arm might be material stripped from galaxies in this chain. The origins of arm A4 are uncertain. Away from the core of the cluster there are some other velocity subgroups in that direction (*5, 19*), but alternatively A4 could consist of material stripped from galaxies associated with a NGC 4889 subgroup, because the arm ends at that galaxy.

The cluster contains a well-studied giant radio halo (*20*). Such emission is exclusively found in merging clusters (*21*), but the source of the relativistic electrons required is still a matter of debate. The electrons could be accelerated by shocks or turbulence in the ICM, or they could be generated by relativistic proton collisions in the cluster. If we compare the observed X-ray structures with previous low frequency radio maps (*22*), we do not see any features in common (Fig. S3). Therefore, we do not observe emission from electron acceleration processes, for example turbulence or shocks, associated with the arms. The outer edge of the L1 feature coincides with the edge of the radio emission and with a shock detected using Planck and XMM (*23*).

## Acknowledgments

ACF acknowledges the support of the Royal Society. We thank S. D. Brown for providing the 352 MHz radio map. Support for this work was provided by the National Aeronautics and Space Administration through Chandra Award Number GO2-13145X issued by the Chandra X-ray Observatory Center, which is operated by the Smithsonian Astrophysical Observatory for and on behalf of the National Aeronautics Space Administration under contract NAS8-03060. The scientific results reported in this article are based on observations made by the Chandra X-ray Observatory data obtained from observations made by the Chandra X-ray Observatory and data obtained from the Chandra Data Archive. Based on observations obtained with XMM-Newton, an ESA science mission with instruments and contributions directly funded by ESA Member States and NASA. The data analyzed here can be found in the XMM-Newton Science Archive and Chandra Data Archive.

# Supplementary Materials

## Materials and Methods

We assume that $H_0 = 70$ km s$^{-1}$ Mpc$^{-1}$. At the distance of the Coma cluster, $z = 0.0231$ (*24*), 1 arcsec on the sky is 0.47 kpc. All coordinates use the J2000 system. North is to the top and east is to the right in all images. Any uncertainties quoted are 1σ.

## 1 Chandra data analysis

We observed the Coma cluster with Chandra for a total of 479 ks, pointing at the high density region between the central galaxies. We also incorporated 67 ks of existing archival data into the analysis. The Chandra datasets we analyzed are listed in Table S1. We created light curves for the ACIS-I detector in the 0.5 to 12 keV band using 200s time bins for each of the observations (Fig. S4). We examined these lightcurves by eye and found no flares. We therefore did not filter any time periods from the data. The datasets were reprocessed using the standard CIAO software package (*25*) tool ACIS_PROCESS_EVENTS, applying VFAINT processing, for those datasets taken in this mode, to reduce the non X-ray background. These reprocessed event files were then filtered to only include standard grades. We next reprojected the data to the common coordinate system of observation 13994 using REPROJECT_EVENTS. Images of the reprojected datasets were created (using the standard ACIS pixel size) and added together to make a merged image.

For each CCD in each observation we created exposure maps using the MKEXPMAP tool in the 0.5 to 7 keV band. As the exposure map is energy dependent, we input a model spectrum to MKEXPMAP, with a temperature of 9.7 keV, metallicity of 0.24 Solar and absorbing column density of $2.5 \times 10^{20}$ cm$^{-2}$. The exposure maps were added together to make a total exposure map. After merging, it became clear that there were some artifacts in the exposure-corrected images associated with the edges of the ACIS-I detector. To remove these, we used a square region filter to exclude the detector edges in the images and exposure maps. The total exposure map of the central region is shown in Fig. S1 (right).

We used standard blank sky background files to remove the X-ray background from the images and during spectral fitting. For each input observation file we looked up the appropriate background event files for the ACIS-I CCDs with ACIS_BKGRND_LOOKUP. These files were merged together with DMMERGE. The merged event file was reprojected to match the observation file, using the appropriate aspect file for the observation. Because the background files did not have the same bad pixel maps as the observations, we filtered events from the backgrounds which occurred in bad pixels in the foregrounds. We then reprojected the background a second time to the common 13994 coordinate system. For those observations which were taken in VFAINT mode, we applied VFAINT filtering to the background event file. Because of the change in ACIS background over time, we adjusted the exposure time of the background observation so that the rates in the 9 to 12 keV band (which is dominated by particle background) matched the foreground observation.

We created images from each of the background observations, scaling to match the exposure times of their respective foreground images. The background images were added to make a total background image. We subtracted this background from the mosaiced foreground and divided by the total exposure map, to make the final exposure-corrected image.

To identify point sources in the images, we applied the CIAO WAVDETECT source detection software to the total foreground image in the 0.5 to 7 keV band with the combined exposure map. An average PSF map for all the observations was used in the source detection. To make this map, PSF maps were created for each reprojected observation using MKPSFMAP. These were averaged, weighting by the exposure map of each observation. We used wavelet scales of 1, 2, 4 and 8 for the point source detection. We did not use larger scales because the extended cluster emission began to be detected. The point sources found were masked out in subsequent analyses.

Fig. 1A is an exposure-corrected background-subtracted X-ray image in the 0.5 to 7 keV band, smoothed by a Gaussian of σ = 2 arcsec. The unsharp masked image (Fig. 1B) was created by smoothing the 0.5 to 7 keV X-ray image by Gaussian kernels with σ of 11.8 and 63 arcsec and taking the fractional difference of the first image from the second.

Fig. 2 uses different smoothing scales compared to Fig. 1B for the unsharp masking. In this case we subtracting the data smoothed by a Gaussian with σ = 128 pixels (63 arcsec) from the data smoothed by 6 pixels (3 arcsec). Fig. S1 (left) was created similarly using smoothing scales of 63 and 4 arcsec.

## 2  Spectral map creation

The temperature map (Fig. 3) was made by fitting spectra extracted from spatial regions. The regions were chosen using the Contour Binning algorithm (*26*), which creates regions which are designed to follow contours in an image. We used an unsharp masked image for this purpose, which was the difference between an X-ray image smoothed by a Gaussian of 48 0.492 arcsec pixels from an image smoothed by 128 pixels. A geometric constraint factor of 2 was used to prevent the bins from becoming too elongated. The bins were chosen to grow to a signal to noise ratio of 200, giving each bin around $4\times10^5$ counts.

Spectra were extracted from each of the regions from each of the observations. Response and ancillary response matrices were created for each observation, using MKACISRMF and MKWARF, respectively, weighting according to the number of counts in the 0.5 to 7 keV band. Background spectra were extracted from the background datasets. We added the foreground spectra from each of the datasets together. The responses and ancillary responses were averaged, weighting according to the relative exposure times of the observations. We also added background spectra. As the ratio of the background to foreground exposures was different for each observation, we shortened the exposure of individual backgrounds before adding until they had the same ratio, by discarding photons from the spectra.

The spectra were fitted using an APEC spectral model (*27*), fitting for the temperature, metallicity and normalization. We used the Solar relative abundances of (*28*). We applied a PHABS photoelectric absorber (*29*) to the model, with a fixed column density of $2.5\times10^{20}$ cm$^{-2}$, chosen to be the average best fitting value when it was unconstrained. We used XSPEC (*30*) to fit the spectra, minimizing the modified C-statistic. The spectra were fit between 0.5 and 7 keV.

An image containing the best fitting temperature values in each bin region was created. This image was rebinned by a factor of 4 to have 1.968 arcsec pixels and finally smoothed by a Gaussian of width σ = 4 pixels, creating the map shown in Fig. 3. The uncertainties in the temperature range from 0.2 keV in the coolest regions to 0.5 keV where hottest. The contours in Fig. 3 were created using an unsharp-masked surface brightness map, taking the fractional difference between the data smoothed by a Gaussian with σ = 15 from that smoothed using 63 arcsec.

## 3 Thermodynamics of the arms

### 3.1 Using the surface brightness to estimate the density increase

To calculate the density of a cylindrical arm embedded in the larger cluster, we assume that the surface brightness is proportional to the density-squared, integrated along the line of sight. At Coma temperatures, the effect of temperature and metallicity on surface brightness are small. The ratio of the density in the arm to the cluster is given by

$$\frac{n_A}{n_C} = \sqrt{\frac{S_A}{S_C}\left(\frac{d_C}{d_A}+1\right) - \frac{d_C}{d_A}},$$

where the surface brightness of the arm is $S_A$, that in the neighboring cluster region is $S_C$, the intrinsic electron density of the arm and cluster are $n_A$ and $n_C$, respectively, and the depth of the arm and cluster are $d_A$ and $d_C$, respectively. To calculate the density we assume that the effective depth of the cluster is the twice the radius from the cluster center. For the cylindrical case, the arm depth is its width on the sky. For the mid-point of A1, the ratio between the depth of the cluster and of the arm is around 8 to 1. A 10% increase in surface brightness implies a 40% rise in density.

The value of 10% surface brightness enhancement for A1 is subject to some uncertainties, because the underlying surface brightness is not completely known. If a larger upper smoothing scale is used for the unsharp masking, values of 15 to 20% can be obtained for the excess emission.

A better way to estimate the density increase might be to assume a profile for the cluster. We fit a β surface brightness model (*31*) to the image, excluding regions, such as the arms, where the surface brightness is enhanced. Using this model we obtain an increase in surface brightness in the arms of around 15% relative to the model. The β fit has a core radius of 7.4 arcmin and a β parameter of 1.1. Examining XMM data, (*15*) found parameters of 9 arcmin and 0.9, respectively, which do not significantly affect our calculations.

The *β* model can be trivially deprojected to a 3D emissivity or density model. The emissivity can then be integrated along the line of sight and the surface brightness computed, in the presence or absence of an arm in the cluster. Assuming that the arm has a particular fractional density contrast relative to the cluster, we find a 15% surface brightness enhancement in the arm is explained by a density increase of 35%, if the arm lies along the central plane of the cluster, perpendicular to the line of sight.

### 3.2 Spectral fitting when accounting for projection

To examine the pressure of the arms we need to know their intrinsic temperature and density, accounting for projection effects. To do this, we examined regions lying along the two arms A1 and A3 (Fig. S5). We also made neighboring comparison regions, chosen to be close to the arm region with similar underlying surface brightness. Spectra were extracted from the arms and comparison regions, using the method described in Section 2. Responses, ancillary responses and background spectra were also made with the method above.

For each of the arms, we simultaneously fit the spectra from the arms and comparison regions using XSPEC. The spectra were fit between 0.5 and 7 keV, minimizing the modified C-statistic. For the comparison region we used a single APEC thermal component, with PHABS absorption. For the arm region we included an absorbed APEC component for the arm emission itself, but added the comparison region component, scaling by the small differences in area on the sky, and multiplying by a 7/8 factor to account for the reduction in volume because the arm is not present in the comparison region (the arm is assumed to be 1/8 the depth of the cluster). In the spectral fits the temperature, absorption, metallicity and normalization of the components were allowed to be free.

The Markov Chain Monte Carlo (MCMC) technique can be used to examine the uncertainty of model parameters when applied to data (the posterior probability). A set of parameter value steps (the chain) is produced by a sampler from the previous step in the chain. The parameter uncertainties are obtained from the distribution of parameter values within the chain. We used the MCMC functionality in XSPEC with the Metropolis-Hastings sampler to generate the chain. In our analysis, we used a chain length of 10 000, starting the chain with the best fitting parameter values. The final chain acceptance rates were between 20 and 25%, close to recommended values (*32*). We manually inspected the chain to check that it had rapid mixing.

The ratio of the temperatures, densities, pressures and entropies between arms A1 and A3 and their surroundings are shown in Table S2. The temperature and metallicity were directly taken from the mean and standard deviation model parameters in the chain. The density was calculated from the normalization, which is proportional to the density-squared and the volume of the emitting region. For A1, the emission measures imply density enhancements of (40±4)%, agreeing with the surface brightness results. A3 shows a value of (54±3)%.

The pressure was calculated by multiplying the temperature and density for each step in the chain. We also compute a quantity, $S = k_B T n_e^{-2/3}$, related to the usual thermodynamic entropy, where $T$ is the temperature and $n_e$ is the electron density, is often used and called entropy in X-ray astronomy (*33*). It is a useful quantity because it is raised and lowered by heating and cooling processes, respectively, in the intracluster medium.

We assumed that the arm depth was 1/8 of the comparison region. For A1 we also show pressure values assuming that the arm has 1/6 and 1/10 of the depth. Due to the geometrical uncertainties, we cannot distinguish between pressure equilibrium of the arms with the surroundings and mild over-pressuring of the arms.

In the paper we estimate the timescale for the arms to be accelerated from rest to the velocity of the intracluster medium ($v_{icm}$). This depends on the density ratio of the arm to the ICM ($f$) and the radius ($r$), if the arms are cylinders, as $f\pi r/v_{icm}$. The timescale for the energy in the arms to be replaced by conduction can also be estimated. We take into account that the conduction may be unsaturated or saturated (*34*), assume Spitzer conductivity, the ratio of the specific heats, $\gamma$, is 5/3, the length of the temperature gradient is the arm radius (25 kpc) and that there is a 2 keV temperature between the surroundings (at 9 keV) and the arm.

## 4  XMM data processing

The XMM datasets listed in Table S3 were included in this analysis. We ran the XMM Science Analysis System EMCHAIN program to create reduced event files for the MOS detectors. We did not examine the PN data because the PN background is substantially greater than the MOS background and Coma has a relatively low surface brightness. Light curves in 100s bins for the event files were created in the 10 to 15 keV band, using events with a PATTERN of 0. A sigma clipping algorithm was applied to select time periods without flares. The exposure times of the observations before and after filtering are shown in Table S3. The event files were filtered to exclude these time periods, events with PATTERN values greater than 12 and events without a FLAG value of 0.

The event files were reprojected to have a common coordinate system of the 0124711401 observation. Images for each dataset were made in the 0.5 to 4 keV band using a bin size of 25 pixels. Exposure maps were created with using a PI range of 800 to 1600. All the images were added to make a total image. The exposure maps were similarly added together. By dividing the two we created an exposure-corrected image.

To create the fractional difference map in Fig 3, we used a centre of RA 12:59:45.4 and Dec +27:57:32, although the exact position is not important. The 0.5 to 4 keV mosaic was smoothed using a Gaussian of σ = 10 arcsec, before the radial average was calculated and subtracted. Point sources were excluded when calculating the average and in the map.

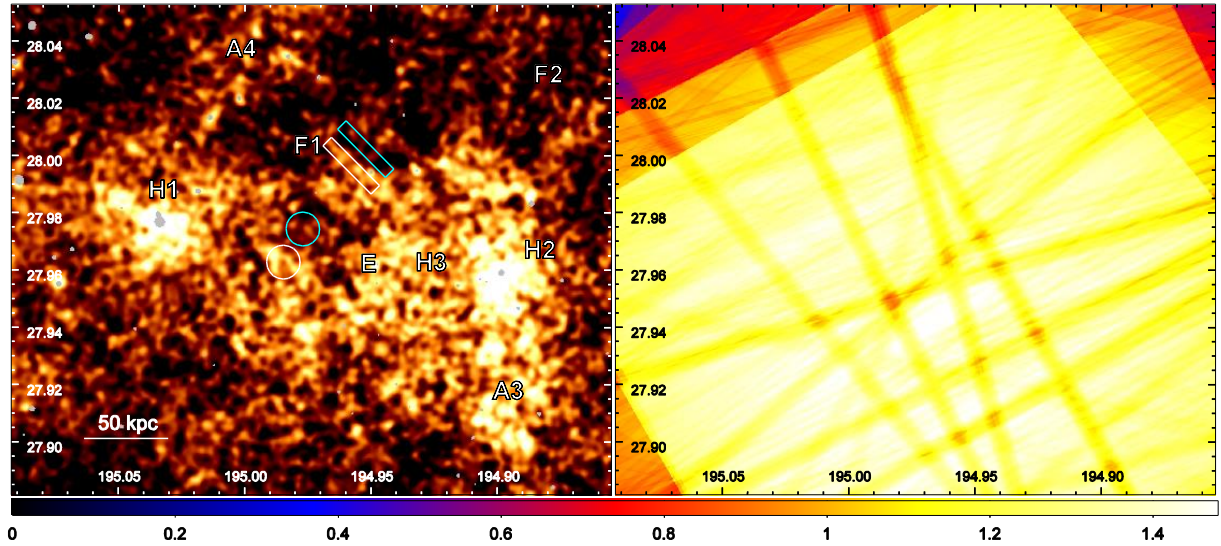

**Fig. S1.** (Left) Chandra unsharp-masked detailed image of the region around the central galaxies (see Fig. 2), with the features labeled. We examined the significance of the features marked by the white rectangle and circle, relative to the cyan comparison regions. The features are detected relative to the comparison regions by 6.8 and 6.0σ, respectively. (Right) Exposure map of the same region, with units of $10^8$ s cm$^2$ counts photon$^{-1}$, shown in the color bar. The filamentary features are not associated with features in the exposure map.

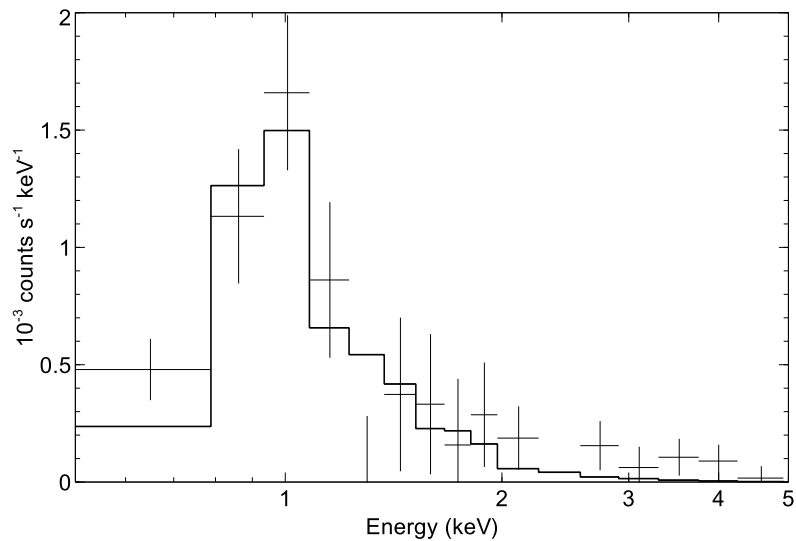

**Fig. S2.** X-ray spectrum of the tail of the galaxy GMP 2910 (labeled T in Fig. 1). We subtracted the cluster emission, taken from neighboring regions (Fig. S5). The data have had the spectral channels binned by a factor of 10 for display. The solid line shows the best fitting APEC spectral model with Solar metallicity and absorption of $2.5\times10^{20}$ cm$^{-2}$, found by minimizing the modified C-statistic in XSPEC. The best fitting temperature is 1.0±0.1 keV and is insensitive to the metallicity, providing it is above 0.1 Solar. The same temperature is obtained if the cluster emission is jointly fit in the tail and neighboring regions, with an additional component for the tail emission.

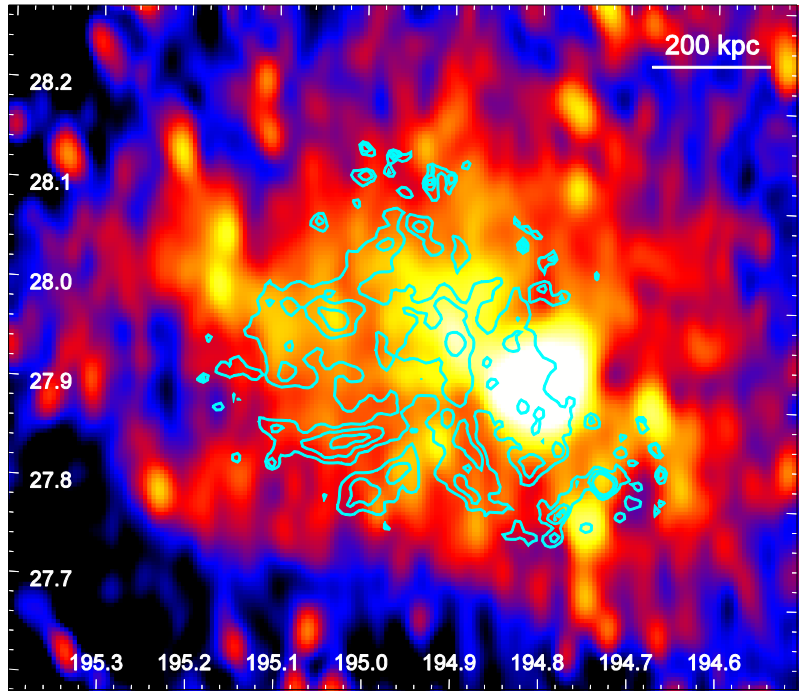

**Fig. S3.** Comparison of the WRST 352 MHz radio emission (*22*) with Chandra excess emission contours (see Figs. 3 and 4). Note that much of the bright emission to the east of the centre is due to the radio galaxy NGC 4874 and the neighboring tailed-source NGC 4869.

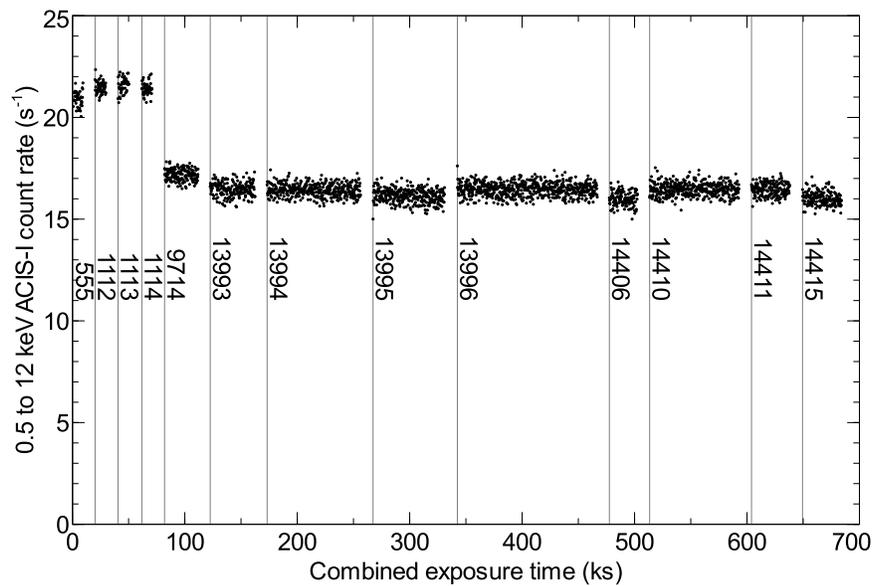

**Fig. S4.** Combined light curves of the Chandra observations. The light curves have been plotted on the same horizontal axis, separated by gaps of 10 ks. We label the segments of the light curve by the observation identifier. The light curves were created for the ACIS-I CCDs in 200 s bins in the 0.5 to 12 keV band. Note that the average count rates vary between the observations due to several factors, including changes in effective area and background as a function of time, the difference in background rate between VFAINT and FAINT detector modes and the different pointing directions and roll angles.

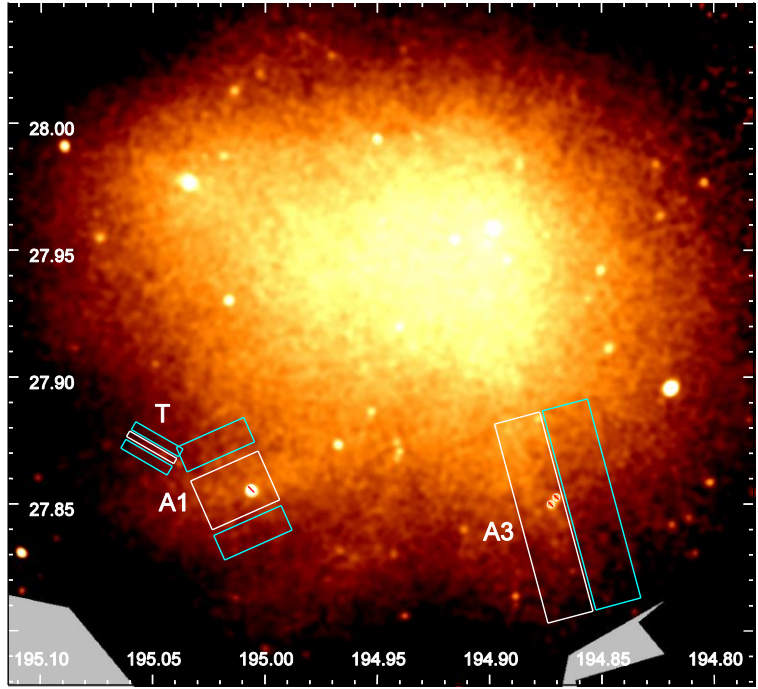

**Fig. S5.** Spectral extraction regions. Shown as white boxes are the regions used for spectral analysis for arms A1 and A3, and the tail of the galaxy GMP 2910 (T). Comparison spectra were extracted from the cyan boxes (two for A1 and T, and one for A3). The circles are excluded point sources.

**Table S1.** List of Chandra datasets analyzed in this paper. We show the observation identifier, date of the start of the observation, detector observing mode and the total exposure time.

| Observation | Date | Mode | Exposure (ks) |
| --- | --- | --- | --- |
| 555 | 1999-11-04 | FAINT | 8.7 |
| 1112 | 1999-11-04 | FAINT | 9.7 |
| 1113 | 1999-11-04 | FAINT | 9.6 |
| 1114 | 1999-11-04 | FAINT | 9.0 |
| 9714 | 2008-03-20 | VFAINT | 29.6 |
| 13993 | 2012-03-21 | VFAINT | 39.6 |
| 13994 | 2012-03-19 | VFAINT | 82.0 |
| 13995 | 2012-03-14 | VFAINT | 63.0 |
| 13996 | 2012-03-27 | VFAINT | 123.1 |
| 14406 | 2012-03-15 | VFAINT | 24.8 |
| 14410 | 2012-03-22 | VFAINT | 78.5 |
| 14411 | 2012-03-20 | VFAINT | 33.6 |
| 14415 | 2012-04-13 | VFAINT | 34.5 |

**Table S2.** Ratios of quantities between the arm and the surroundings, obtained by spectral fitting. A depth for the arms of 1/8 of the cluster is typically assumed. The ratio was calculated at each step within the chain. The values shown are the median and the range containing 68.2% of the steps.

| Arm | Depth | Quantity | Ratio |
|---|---|---|---|
| A1 | 1/8 | Temperature | 0.85±0.10 (= 7.3 keV/ 8.6 keV) |
| | | Density | 1.40±0.04 |
| | | Pressure | 1.19±0.14 |
| | | Entropy | 0.67±0.08 |
| | | Metallicity | $0.9^{+0.7}_{-0.6}$ |
| | 1/6 | Pressure | $1.29^{+0.14}_{-0.12}$ |
| | 1/10 | Pressure | 1.10±0.13 |
| A3 | 1/8 | Temperature | 0.81±0.07 (= 7.3 keV / 9.0 keV) |
| | | Density | 1.54±0.03 |
| | | Pressure | 1.24±0.10 |
| | | Entropy | 0.61±0.05 |
| | | Metallicity | $1.0^{+0.7}_{-0.5}$ |

**Table S3.** List of XMM datasets examined in this paper. The exposure times show the mean exposure time of the MOS detectors before filtering for flares (initial) and afterwards (filtered).

| Observation | Date | Initial exposure (ks) | Filtered exposure (ks) |
|---|---|---|---|
| 0124710301 | 2000-06-27 | 26.5 | 15.4 |
| 0124710401 | 2000-06-23 | 18.9 | 11.5 |
| 0124710501 | 2000-05-29 | 25.8 | 24.1 |
| 0124710601 | 2000-06-12 | 18.0 | 10.1 |
| 0124710701 | 2000-06-24 | 11.7 | 5.8 |
| 0124710801 | 2000-12-10 | 27.4 | 24.2 |
| 0124710901 | 2000-06-11 | 29.9 | 21.1 |
| 0124711101 | 2000-06-24 | 35.8 | 24.2 |
| 0124711401 | 2000-05-29 | 20.3 | 16.5 |
| 0124712001 | 2000-12-10 | 21.8 | 15.8 |
| 0124712101 | 2000-12-10 | 27.0 | 26.1 |
| 0124712201 | 2000-12-09 | 26.5 | 25.8 |
| 0124712401 | 2002-06-05 | 26.6 | 18.2 |
| 0124712501 | 2002-06-07 | 27.7 | 27.0 |
| 0153750101 | 2001-12-04 | 25.0 | 20.7 |
| 0204040101 | 2004-06-06 | 86.5 | 72.0 |
| 0300530101 | 2005-06-18 | 25.0 | 21.4 |
| 0300530201 | 2005-06-17 | 8.7 | 7.7 |
| 0300530301 | 2005-06-11 | 30.4 | 29.9 |
| 0300530401 | 2005-06-09 | 26.9 | 18.9 |
| 0300530501 | 2005-06-08 | 24.9 | 24.4 |
| 0300530601 | 2005-06-07 | 24.9 | 22.1 |
| 0300530701 | 2005-06-06 | 25.0 | 23.8 |